\documentclass[12pt]{article}
\usepackage{epsfig}
\newcommand{\noi}{\noindent}
\newcommand{\eq}{\begin{equation}}
\newcommand{\en}{\end{equation}}
\newcommand{\eqa}{\begin{eqnarray}}
\newcommand{\ena}{\end{eqnarray}}

\newcommand{\cao}{{\cal O}}

\newcommand{\tr}{\mbox{Tr}}

\newcommand{\aleq}{\mbox{}_{\textstyle \sim}^{\textstyle < }}

\begin{document}

\renewcommand{\baselinestretch}{1.1}
\renewcommand{\thefootnote}{\arabic{footnote}}
\setcounter{footnote}{0}
\language0

\hbox{}
\noindent October 2001

\vspace{0.3cm}
\begin{center}

\renewcommand{\thefootnote}{\fnsymbol{footnote}}
\setcounter{footnote}{1}

{\large PJLZ--gauge fixing approach in $SU(2)$ lattice gauge theory.}
\footnote{Supported by the grant INTAS-00-00111 and the RFRB 
grants 99-01-01230 and 01-02-17456}

\vspace*{0.5cm}
V.K.~Mitrjushkin$\mbox{}^{a,b}$ and A.I. Veselov$\mbox{}^b$

\end{center}

\vspace*{0.3cm}
{\sl \noindent
 \hspace*{6mm} $^a$ Joint Institute for Nuclear Research, 141980 Dubna, 
          Russia \\
 \hspace*{6mm} $^b$ Institut of Theoretical and Experimental Physics,
                    Moscow, Russia}

\vspace{0.3cm}

\begin{center}
{\bf Abstract}
\end{center}

{\small
We study the $SU(2)$ gauge theory with the
interpolating gauge {\it a la} Parrinello--Jona-Lasinio--Zwanziger
(PJLZ) with the gauge fixing functional $F=\sum_{x\mu} \frac{1}{2}\tr
(U_{x\mu}\sigma_3U_{x\mu}^{\dagger}\sigma_3)$.
We find a strong indication of the non--analiticity with respect to
the interpolating parameter $\lambda$ at $\lambda_c \sim 0.8$.
}
\vspace{0.5cm}

\section{Introduction} 
\setcounter{equation}{0}

Gauge variant objects, i.e. Green functions for gluons or/and quarks, 
are among the most popular objects in continuum physics. A comparison 
of nonperturbatively calculated Green functions on the lattice with 
continuum (mainly, perturbative) ones can give an insight on the 
structure of the lattice theories and role of nonperturbative effects. 
Another important point is that Green functions are supposed to contain 
information about the physical `observables' which must not depend on 
the gauge chosen, e.g. dynamical gluon masses, screening masses, etc.. 
Therefore, it is important to disentangle gauge--dependent features 
from gauge independent ones.

A somewhat special reason to study the gauge, interpolating between
no--gauge and Maximally Abelian gauge (MAG), is connected with 
a fate of the so called Abelian Dominance.
Recently Ogilvie has shown \cite{ogil} (see also \cite{fgro}) that for 
Abelian Projection (AP) the gauge fixing is unnecessary, i.e. AP 
without gauge fixing yields the exact string tension of the underlying 
non--Abelian theory :  $\sigma_{Abel} = \sigma_{SU(2)}$.

These observations shed a new light on the problem of Abelian 
Dominance. Indeed, without MAG the Abelian Projection ensures the exact 
equality between $\sigma_{Abel}$ and $\sigma_{SU(2)}$ while with MAG 
Abelian $\sigma_{Abel}$ and full $\sigma_{SU(2)}$ string tensions are 
close but not equal :  $\sigma_{Abel}\ne \sigma_{SU(2)}$, at least for 
$\beta$--values employed (see, e.g.  \cite{stac,bkpv}). 

The question arises if it is possible to interpolate `smoothly' from 
the no--gauge case to the gauge fixed case. 
To answer this question is the main goal of this work.

\section{Gauge fixing procedure and algorithm} 
\setcounter{equation}{0}

We consider the pure gauge $SU(2)$ theory with standard Wilson action 
$\beta \cdot S(U)$ \cite{wils}.
According to PJLZ--approach \cite{pajl},\cite{zwan} the average of any 
gauge--noninvariant functional $~\cao(U)~$ is given by

\eq
\langle \cao \rangle = \frac{1}{Z}\int\! [dU]
~ {\widetilde \cao}(U;\lambda) \cdot e^{-\beta S(U)}~,
\label{pjlz0}
\en

\noi where ${\widetilde \cao}(U;\lambda)=\langle \cao\rangle_{\Omega}$
and

\eqa
\langle \cao\rangle_{\Omega}
&=& \frac{1}{I(U;\lambda)}\int\! [d\Omega]
~\cao(U^{\Omega}) \cdot e^{\lambda F(U^{\Omega})} ;
\label{pjlz}
\\
I(U;\lambda) &=& \int\! [d\Omega]~ e^{\lambda F(U^{\Omega})}~,
\nonumber
\ena

\noi where $F(U)$ is a gauge fixing functional and
$U_{x\mu}^{\Omega}=\Omega_x U_{x\mu} \Omega_{x+\mu}^{\dagger}$.
We have chosen 

\eq
F = \sum_{x\mu} 
\frac{1}{2}\tr (U_{x\mu}\sigma_3U_{x\mu}^{\dagger}\sigma_3).
\label{func1}
\en

\noi In eq.(\ref{pjlz}) the functional $F_U(\Omega)$ plays the role of 
effective action with unitary `spins' $\Omega_x$ and random bonds 
described by fields $U_{x\mu}$ (similar to spin--glass model).

Evidently, the maximization of $F(U^{\Omega})$ with respect to gauge 
transformations $\Omega$ defines MAG, and for gauge invariant 
functional ${\widetilde \cao}(U;\lambda)=\cao(U;\lambda)$.  

In eq.'s (\ref{pjlz}) $~\lambda$ is some 
`interpolating' parameter between $0$ and $\infty$. The choice 
$\lambda=0$ corresponds to the no--gauge case and the 
limit $\lambda\to\infty$ corresponds to the case of the Maximally 
Abelian gauge. Any physical, i.e. gauge invariant, observable
(screening masses, etc.) must not depend on $\lambda$.
In general, there is no grounds for saying that one value 
of $\lambda$ is more physical than another value. However, things can 
be different in the case of the Abelian Projection if $\lambda=0$ and 
$\lambda=\infty$ belong to two different phases. 

In our study we use $\cao=F(U)$ defined in eq.(\ref{func1}) and 
$F_{norm}(U)=F(U)/4V_4$.  In the `strong coupling' approximation 
($\lambda \sim 0$) one obtains

\eq
\langle F_{norm}\rangle_{str} = \lambda/3 + \ldots~,
\label{strong1}
\en

\noi where $V_4$ is the number of sites. 

Definitions in eq.'s (\ref{pjlz0})$\div$(\ref{pjlz})  
presume the following numerical algorithm \cite{hopr}.  

\begin{itemize} 

\item[i)] Generate a set of link configurations $\{U_{x\mu}^{(1)}\}, 
\{U_{x\mu}^{(2)}\}, \ldots$ using standard gauge invariant algorithm 
with Wilson action $S(U)$ at some value of $\beta$.  

\item[ii)] For every configuration $\{U_{x\mu}^{(i)}\}$ generate
sequence of configurations $\{\Omega_x^{(1)}\}$, $\{\Omega_x^{(2)}\}, 
\ldots$ weighted by the factor $\exp(\lambda F(U^{\Omega}))$ at some 
value of $\lambda$. Therefore, one obtains the estimator for
${\widetilde \cao}(U;\lambda) = \langle \cao\rangle_{\Omega}$

\eq
{\widetilde \cao}(U;\lambda) = \frac{1}{N_{\Omega}}
\sum_{j=1}^{N_{\Omega}} \cao (U^{\Omega^{(j)}})~.
\en

\item[iii)] The estimator for the expectation value 
$\langle \cao\rangle$ is obtained as 

\eq
\langle \cao\rangle = \frac{1}{N_U} \sum_{i=1}^{N_U}
{\widetilde \cao}(U^{(i)};\lambda)~.  
\en
\end{itemize}

\noi

\section{Numerical results} 
\setcounter{equation}{0}

The most part of our calculations has been performed on the $8^4$ 
lattice at $\beta=2.4$. Some calculations have been done also on the 
$6^4$ and $10^4$ lattices to control finite volume effects.


\begin{figure}[tbp]
\vspace{-1cm}
\centerline{
\epsfxsize=9cm\epsfysize=9cm\epsffile{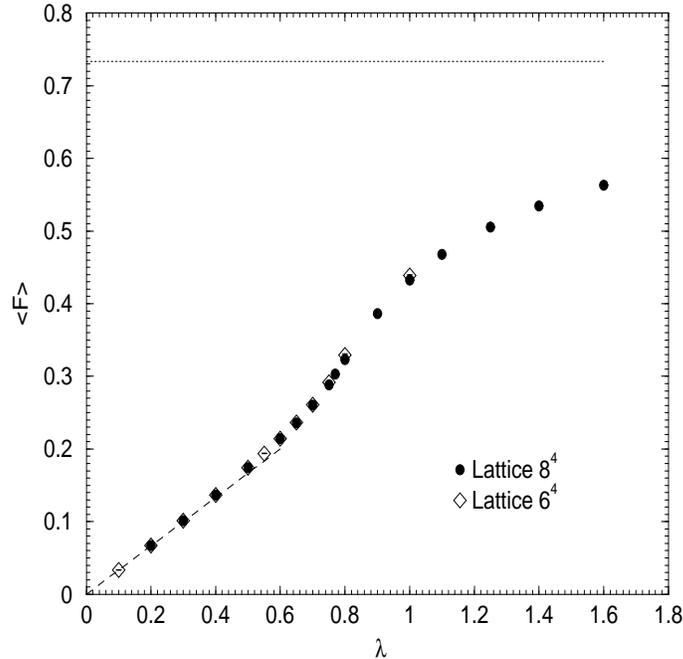}
}
\vspace{-0.5cm}
\caption{The dependence of the $\langle F_{norm}\rangle$ on 
$\lambda$.  Symbols are explained in the text.
} 
\label{fig:fu} 
\end{figure} 
\begin{figure}[tbp]
\vspace{-1cm}
\centerline{
\epsfxsize=9cm\epsfysize=9cm\epsffile{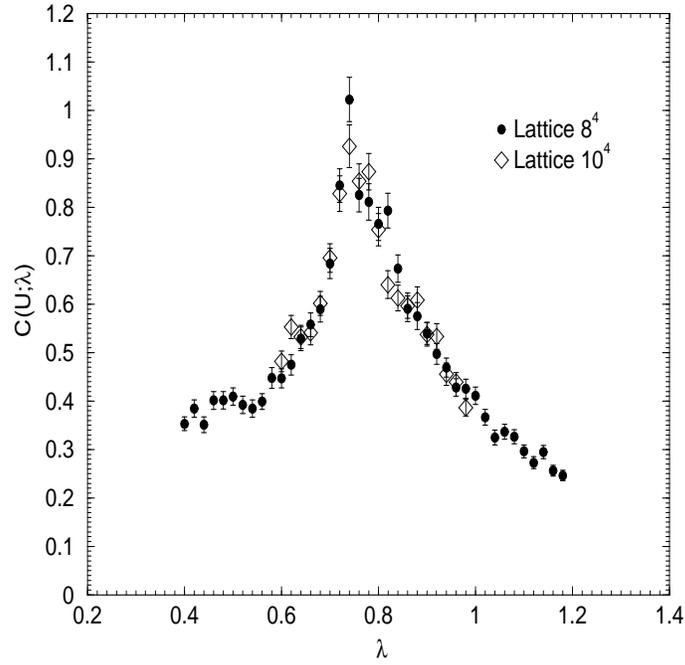}
}
\vspace{-0.5cm}
\caption{The dependence of $C(U;\lambda)$ on 
$\lambda$ for some typical configuration $\{U_{x\mu}\}$.
} 
\label{fig:cu} 
\end{figure} 
\begin{figure}[tbp]
\vspace{-1cm}
\centerline{
\epsfxsize=9cm\epsfysize=9cm\epsffile{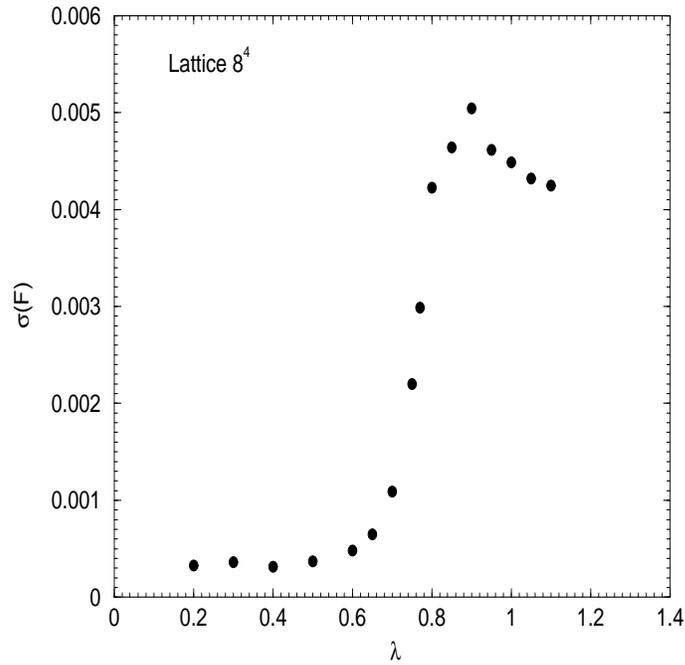}
}
\vspace{-0.5cm}
\caption{The dependence of $\sigma(F)$ on $\lambda$.
} 
\label{fig:sigm1_fu} 
\end{figure} 

In Figure \ref{fig:fu} one can see the dependence of the 
$\langle F_{norm}\rangle$ on $\lambda$ at $\beta=2.4$. The 
dashed line correspondes to the lowest order `strong coupling' 
approximation $\langle F\rangle_{strong} = \lambda/3$.  The upper 
dotted line corresponds to Maximally Abelian gauge.
The agreement between numerical data and `strong coupling' expansion in 
eq.(\ref{strong1}) is very good up to $\lambda \simeq 0.6$. 
$\langle F_{norm}\rangle$ has a clear change of regime at $\lambda \sim 
0.8$. It is interesting to note that this dependence is very similar 
to that for another choice of the functional $F=F_{LG}$ \cite{hopr}
which corresponds to the Lorentz (or Landau) gauge at
infinite values of the interpolating parameter.

For any $~\{U_{x\mu}\}$--configuration 
the `specific heat' $~C(U;\lambda)~$ is defined as

\eq
C_U(\lambda) = \frac{1}{4V_4}
 \frac{d {\widetilde F}(U;\lambda)}{d\lambda}
= \frac{ \langle F^2 \rangle_{\Omega}
- \langle F \rangle_{\Omega}^2}{4V_4} .
\en

\noi In Figure \ref{fig:cu} we show the dependence of $C_U(\lambda)$
on $\lambda$ for some typical configuration $\{U_{x\mu}\}$.
One can see a sharp propounced peak (cusp) at $\lambda_c \sim 0.8$. Of 
course, the position and size of this peak depend somewhat on the 
choice of configuration. However, this peak demonstrates rather weak 
dependence on the volume (compare $8^4$ and $10^4$ data).

Let us define the variance $\sigma(F)$ in a standard way

\eq
\sigma^2(F) = \frac{1}{N_U}\sum_i^{N_U} {\widetilde F}^2_i
- \left( \frac{1}{N_U}\sum_i^{N_U} {\widetilde F}_i \right)^2~.
\en

\noi Figure \ref{fig:sigm1_fu} demonstrates the dependence
$\sigma(F)$ on $\lambda$ for $8^4$ lattice. For comparatively small 
values of $\lambda$, i.e. till values $\lambda \aleq 0.6$ where the 
strong coupling approximationfor $\langle F\rangle$ works well, this 
variance is practically stable. 
However, for $\lambda$'s between $0.65$ and $0.8$ nee can see a drastic 
increase of the variance.

\section{Summary and discussion}

To summarize, we have performed an exploratory study of the pure gauge 
$SU(2)$ theory with the interpolating gauge {\it a la} 
Parrinello--Jona-Lasinio--Zwanziger with the gauge fixing functional 
defined in eq.(\ref{func1}). Therefore, this gauge interpolates between 
no--gauge case and maximally Abelian gauge.

Our data indicate on the existence of the strong non--analyticity with
respect to $\lambda$ (phase transition) at $~\lambda_c \sim 0.8$.
Most probably, the mechanism of this transition is similar to that in
the spin--glass models.  At the moment it is rather difficult to
specify the order of this phase transition.
It is interesting to note that the existence of a transition with
respect to the interpolating parameter $\lambda$ has been also found
for another choice of the functional $F=F_{LG}$ \cite{hopr} which
corresponds to the Lorentz (or Landau) gauge at infinite values of
$\lambda$.

The existence of this transition makes it clear that there is no
smooth interpolation between the no--gauge case and the case with MAG.
This observation is of importance for gauge dependent objects
(e.g. $\sigma_{abel}$) especially taking into account that the Abelian
Projection $~\{U_{x\mu}\} \to \{ h(U_{x\mu})\}~$ is (not very well
controllable) approximation.
We conclude that the `physics' of Abelian Projection is
supposed to be different at $\lambda=0$ (where
$\sigma_{abel}=\sigma_{SU(2)}$) and the case with MAG.

The above conclusion needs further confirmation. Finite volume
effects must be better studied as well as the dependence of other
observables (e.g. $\sigma_{abel}$) on $\lambda$. This work is in
progress.

\end{document}